\documentclass[aps]{revtex4}
\newcommand{\dslash}{D\!\!\!\!\slash}

\begin{document}

\title{QCD Functional Integrals for Systems with Nonzero Chemical
Potential }


\author{Thomas D.~Cohen\,\footnote{\,\uppercase{W}ork supported in part by
the \uppercase{U}.\uppercase{S}. \uppercase{D}epartment of
\uppercase{E}nergy under grant  \uppercase{DE}-
\uppercase{FG}02-93 \uppercase{ER}-40762.}}

\affiliation{Department of Physics, University of Maryland \\
College Park, MD 20742 USA\\
E-mail: cohen@physics.umd.edu}

\begin {abstract} This paper reviews some recent progress on QCD
functional integrals at nonzero chemical potentials.  One issue
discussed is the use of QCD inequalities for this regime.  In
particular, the positivity of the integrand of particular
Euclidean space functional integrals for two-flavor QCD with
degenerate quark masses is used to demonstrate that the free
energy per unit volume for QCD with a baryon chemical potential
$\mu_B$ (and zero isospin chemical potential) is necessarily
greater than the free energy with isospin chemical potential
$\mu_I = \frac{2 \mu_B}{N_c} $ (and zero baryon chemical
potential).  This result may be of use in model finite density
systems.  A corollary to this result is a rigorous {\it ab
initio} bound on the nucleon mass. The second major issue
addressed is the so-called ``Silver Blaze'' problem: the fact
that at zero temperature and chemical potentials less than some
critical value the free energy remains as that of the vacuum.
This is puzzling in the context of a functional integral since a
chemical potential affects the functional determinant of the
Dirac operator and any nonzero $\mu$ changes every eigenvalue of
the Dirac operator compared to the $\mu=0$ value.   The isospin
Silver Blaze problem is solved through the study of the spectrum
of the operator $\gamma_0 (\dslash + m)$. The status of the
baryon Silver Blaze problem is briefly discussed.\end{abstract}

\maketitle
\section{Introduction \label{intro}}

The problem of QCD at nonzero density is important both
phenomenologically and theoretically.  Unfortunately, it is a
problem of formidable difficulty.

There is no known analytical way to attack the problem in terms
of a convergent systematic expansion except at very high
density.  In the very high density regime one can use the fact
the system is weakly coupled to deduce the form of an interaction
kernel between quarks which gives rise to a gap equation yielding
color superconductivity \cite{Son,rho}.  Unfortunately, this
regime is only known to be valid at exponentially high densities.
Accordingly it is doubtful whether this regime is relevant either
in astrophysics or in laboratory experiments.

One might hope to learn about the system via numerical
simulations of lattice QCD\cite{lat}.  Here, too, is a problem.
The standard Monte Carlo algorithm relies on a functional with a
positive definite measure.  Typically finite densities are
achieved via a chemical potential and the chemical potential
generally yields a functional determinant which is not positive
definite and this notorious fermion sign problem spoils the Monte
Carlo approach. One way to avoid this is to concentrate on the
case of QCD with an isospin chemical potential rather than a
baryon one.  This has  the virtue of having a manifestly real and
positive measure in the functional integral \cite{AlfKapWil99}.
Numerical simulations have been done for this
system.\cite{Kog1,Kog2}  Unfortunately, this problem is of little
interest phenomenologically since it is relevant to no known
physical circumstance either in astrophysics or in a doable
terrestrial experiment.  There has been recent progress in ways to
treat systems with finite baryon chemical potentials. However,
these approaches are restricted to the regime of high temperature
and low density.  While this regime is certainly of some interest
and the techniques may be suitable for reaching the QCD critical
point (which is certainly of interest) the techniques are not
suitable for the interesting regime of relative cold matter which
is of real astrophysical interest.

Thus, for much of the regime of interest one is compelled to
resort to model building. Clearly it is of interest to see
whether there are any new analytically  approaches to this class
of problems which can provide new insights.  Such insights may be
of use in furthering theoretical understanding.  They also may
serve to constrain model building.

This paper reports on two related new developments based on
formal properties of the QCD functional integrals at nonzero
chemical potentials.

The first is based on the techniques of QCD
inequalities\cite{LamNus00}. The key insight is that the
functional integral for QCD with a finite baryon chemical
potential differs from QCD with a finite isospin chemical
potential only by a phase.  This in turn lets one bound the free
energy for QCD with a nonzero baryon chemical potential (and zero
isospin chemical potential) by the free energy for QCD with a
nonzero isospin chemical potential (with zero isospin chemical
potential)\cite{TDC1}.  This result is of interest theoretically
and may be of importance in constraining model building---since
the isospin chemical potential case may be simulated on the
lattice, one has a calculable bound which models must not violate
to be consistent with QCD.  The method also has a surprising
spin-off---it provides a rigorous bound on the mass of the
nucleon.

The second development discussed concerns the so-called ``Silver
Blaze'' problem\cite{TDC2}.  This problem is named after the
famous Arthur Conan Doyle story of that name.  In this story
Sherlock Holmes used the ``curious incident'' of a dog doing
nothing in the night time as a key clue.  In the context of QCD
at nonzero chemical potential, the problem arises when one tries
to understand the behavior of QCD at zero temperature and small
chemical potential via the analysis of functional integrals. We
know, of course, that at zero temperature the physical system is
unaffected by a chemical potential which is less than some
critical value. (For the case of an isospin chemical potential
the value is $m_\pi$; for the case of a baryon chemical potential
it is the energy per nucleon of infinite nuclear matter.)  Of
course, from the phenomenological perspective this not surprising
in the least---until one has a chemical potential equal to the
lightest energy per particle number (of the appropriate type) in
the spectrum of the system, then at zero temperature the system
will remain in its vacuum state.  From the point of view of the
functional integral, however, this is a curious incident indeed.
The chemical potential enters the problem through the functional
determinant which is the product of eigenvalues of the Dirac
operator.  The inclusion of {\it any} nonzero chemical potential
alters {\it all} of the eigenvalues. This leads to the natural
expectation that the  nonzero chemical potential will affect all
functional determinants and thus all functional integrals and
thereby all observables. Clearly this does not happen; the
question is simply``why not?''

It can legitimately be asked why one should bother trying to
understand this problem.  This is, quite literally, trying to
understand nothing.  The significance, however, is that if one
wants to ever develop a method based on functional integrals to
understand why something happens when  the critical chemical
potential is exceeded, one has to understand why nothing happens
below. The baryon Silver Blaze problem remains unsolved. However,
the isospin Silver Blaze problem was solved last year providing
new insights into the physics of pion condensation\cite{TDC2}.

A word about notation and language.  For simplicity of
presentation, this paper explicitly discusses the case of QCD with
two degenerate flavors ($u$ and $d$).  Everything goes through
without change if one includes any number of heavy flavors so
long as the chemical potentials associated with these flavors is
zero.  Thus the phrase ``baryon chemical potential'' should be
taken to mean ``the part of the baryon chemical potential
associated with the light nonstrange quarks.''

In the following section the use of QCD inequalities to constrain
QCD at finite baryon chemical potential will be discussed.
Following this there will be a short section exploiting the result
to rigorously constrain the nucleon mass directly from QCD.  The
final section is devoted to the Silver Blaze problem.   The
treatment in sect. \ref{sec:ineq1} and \ref{sec:ineq2} is largely
based on ref.~\cite{TDC1} and the discussion borrows heavily from
that work, while the work in the final section is principally
from ref.~\cite{TDC2}.  The discussion  here, however, is more
expansive and considerably more accessible.

\section{QCD Inequalities For QCD at Nonzero Chemical Potential\label{sec:ineq1}}

\subsection{A Brief Review of QCD Inequalities}

QCD inequalities  are an ideal method to learn {\it some}
qualitative features of QCD {\it in a rigorous way} directly from
the theory. Nussinov\cite{Nus83} developed a precursor to the
approach with a demonstration that bounds could be placed on
various hadronic quantities for a large class of models which were
inspired by QCD. The approach itself emerged shortly thereafter
with the realization by Weingarten \cite{Wei83} and Witten
\cite{Wit83} that analogous bounds could be obtained directly from
QCD itself.  The key tool to deriving these was the Euclidean
space functional integral representations of physical quantities.
The method has an undeniable appeal in that one can deduce
certain qualitative features of QCD from first principles even
while being unable to fully solve the theory.   Of course the
method is quite limited.  One gains qualitative as opposed to
quantitative information, and that being only for particular
observables. The information gleaned from them is important,
however. One role they  serve is simply to supplement the
understanding obtained from lattice simulations. They also give
us an analytic means to understanding some features of QCD which
are both observed in the physical world and which can be seen to
emerge from lattice studies.  As seen here, QCD inequalities can
also provide insight and phenomenologically relevant constraints
for certain properties of QCD that are not tractable on the
lattice using standard Monte Carlo algorithms. The QCD inequality
approach is now more than 20 years old and has been reviewed
recently\cite{LamNus00}.  In this subsection a few relevant
features will be quickly reviewed so the the results are
comprehensible; for more details the reader is directed to see
ref.~\cite{LamNus00}.

Before proceeding it is worth noting that the resulting
inequalities are not derived with full mathematical rigor.  The
results cannot strictly be called theorems. However, by the
standards of physicists they are quite rigorous; they use  only
the most vanilla flavored assumptions typically made by
physicists. The approach implicitly assumes that the QCD exists
as a theory, that it is legitimate to use functional integrals to
compute hadronic quantities from the underlying quantum field
theory, that the standard Wick rotation  to Euclidean space from
Minkowski space is permitted, and the like. But no additional
dynamical assumptions specific to QCD are made.

The key to QCD inequalities is almost trivially simple.  One
begins in the standard way by relating a physical quantity of
interest to a Euclidean functional integral over all possible
gauge field configurations. Now suppose a second interesting
quantity is found whose functional integral has the following
feature: the integrand for the second quantity is greater than or
equal to the integrand of the first quantity for {\it every} gauge
configuration.  If one finds such a pair of quantities,  it is
readily known that the second functional integral is necessarily
bigger than the first. Since the two functional integrals are
related to physical observables, one immediately derives bounds on
the physical quantities.

A bound on the free energy  at fixed baryon chemical potential is
the focus of this section.  In fact, thermodynamically intensive
quantities such as free energy densities are not typically
studied via QCD inequalities. The method is more commonly applied
to the  the study of correlation functions which are then used to
bound the masses of particles.  There is one important example
from the past, however, where the approach used intensive
quantities: the demonstration by Vafa and Witten \cite{VafWit84}
where the vacuum energy for QCD with a $\theta$ term has an
absolute minimum at $\theta=0$.  In fact, this Vafa-Witten paper
\cite{VafWit84} is the subject of some considerable
controversy\cite{questions}.  However, the controversy concerns
the extension of this argument to conclude that parity cannot be
spontaneously broken.   The underlying demonstration that the
vacuum energy has a minimum $\theta=0$ is clearly correct.  The
Vafa-Witten proof will be discussed next as it provides a
template for the bound on the free energy density of QCD at fixed
baryon chemical potential.

The  derivation by Vafa and Witten is both simple and elegant. The
functional integral for the partition function is given by
\begin{equation}
Z(\theta)= \int D[A] \prod_{i=\,{\rm flavors}} \, {\rm det}\big(\dslash
+ m_i\big) \, e^{-S_{YM} \, + \, i \theta \nu }
\end{equation}
where $\, {\rm det}(\dslash + m_i) $ is the functional determinant
for a particular flavor and is known to be both real and
non-negative \cite{Wei83}.  The Euclidean space Yang-Mills action
is denoted by $S_{YM}$, and the topological winding number is
denoted by $\nu$. Consider what happens when one sets $\theta$ to
be nonzero.  The  only effect of doing this is to include a pure
phase factor $e^{i \theta \nu }$ relative to the $\theta=0$ case.
Now the rest of the integrand is real and non-negative and the
real part of this phase factor is always less than or equal to
unity. (We can ignore the imaginary part since we know on
physical grounds that it will integrate to zero.)  Thus, even
without being able to compute the functional integral explicitly
one can deduce that the functional integral for the partition
function with nonzero $\theta$ is smaller than the partition
function with $\theta=0$. But the partition function has a
well-known physical meaning: $Z(\theta) = e^{- V{ E}(\theta)}$
where $V$ is the four-dimensional volume and ${ E}(\theta)$ is the
vacuum energy as a function of $\theta$.  Thus, the bound on the
partition function  implies that ${ E}(\theta) > { E}(0)$.

\subsection{A QCD Inequality for Free Energies at Nonzero Chemical
Potential \label{ssq}}

In this section, a derivation quite analogous to that of Vafa and
Witten discussed above is presented.  As noted in the
introduction, the explicit problem discussed will be  for the  of
two flavor QCD with degenerate quark masses at a nonzero chemical
potential.  As was also noted in the introduction the
generalization to the problem of addition flavors is quite
straightforward.

Like the Witten-Vafa case, the starting point is an appropriate
free energy density.  We begin by considering QCD at fixed
temperature and a  baryon chemical potential, ${ G}_B(T,\mu_B)$.
The free energy is related to the grand partition function
$Z_B(T, \mu_B)$ in the usual way,
\begin{equation}
{ G}_B(T,\mu_B) \, = - \,(\beta V_3)^{-1} \log \big( Z_B(T,
\mu_B) \big ) \;,
\end{equation}
 where $V_3$ is the (three-dimensional) volume of
the system while $\beta$ is the inverse temperature. The next
step is to express this grand partition function  as a functional
integral.  For QCD with two degenerate flavors this is given by
\begin{equation}
Z_B(T,\mu_B) \, = \int  d [A] \, \Big( \, {\rm det}\big(\dslash + m
- \frac{\mu_B}{N_c} \, \gamma_0 \,\big ) \, \Big)^2 e^{-S_{YM}} \;.
\label{fibcp}\end{equation} In the preceding equation $N_c$ is the
number of colors (3 for the physical world), the functional
determinant is taken for one quark flavor.  Temperature is
treated in the standard way via the imposition of boundary
conditions: the gluon fields are subject to periodic boundary
conditions in time $A(t+\beta)=A(t)$ with $\beta=1/T$;  the
fermions in the functional determinant are subject to
antiperiodic boundary conditions. Note that while the functional
determinant is for a single flavor, it comes in squared reflecting
the presence of two flavors in the system.  Finally a notational
issue should be considered.  The chemical potential is for the
baryon number (not for the quark number).  The fact that the
chemical potential is for the baryon number necessitates the
factor of $\frac{1}{N_c}$ seen in Eq.~(\ref{fibcp}).

This functional integral cannot be simulated on the lattice via
standard Monte Carlo methods. The difficulty is, of course, the
fermion sign problem which arises from the functional
determinant.  The key feature about the inclusion of a nonzero
chemical potential is that the functional determinant is, in
general, not necessarily  real or positive. However, from the
perspective of the lattice, what is a major problem from the
perspective of QCD inequalities becomes a major opportunity.   In
particular, it allows one to place an upper bound  on the
partition function:
\begin{equation}
 Z_B(T,\mu_B) \le
\int  d [A] \left | \, {\rm det}\big(\dslash + m - \frac{\mu_B}{N_c}
\, \gamma_0 \big) \right |^2 e^{-S_{YM}} \; \; . \label{partitionineq}
\end{equation}
The inequality seen above is clearly quite analogous to the
Vafa-Witten case in equality and stems from the identical
reason---a phase factor in an otherwise positive definite
integrand will always lower the integral relative to an integral
with a phase factor of unity.

Of course, as written,  inequality (\ref{partitionineq}) is of
little interest.  While the left-hand side has a clear physical
interpretation, the right-hand side at present does not.  The
insight which enables the approach to be fruitful is that the
right-hand side can also be related to a physically meaningful
quantity.  In particular, it is quite straightforward to see that
the right-hand side can easily be related to the free energy
density of QCD with an {\it isospin} chemical potential
\cite{AlfKapWil99}.  An isospin chemical potential term enters
the QCD Lagrangian with the form $\mu_I \, \overline{q} \gamma_0
\frac{\tau_3}{2} q$.  The functional integral for the appropriate
grand partition function $Z_I(T, \mu_I) = \exp \left( {- \beta \,
V_3 \, G}_I(T,\mu_I) \right)$ is
\begin{equation}
Z_I(T,\mu_I) =  \int d [A] \, e^{-S_{YM}} \, {\rm det}\big(\dslash +
m - \frac{\mu_I}{2} \, \gamma_0\big) \, {\rm det}\big(\dslash + m +
\frac{\mu_I}{2}\, \gamma_0\big) \; . \label{ZI}
\end{equation}
The expression has two functional determinants---one for each
flavor---and they have opposite signs in their $\mu_I$ terms which
encodes the fact that up and down quarks have opposite values for
$I_3$.

The next steps involve some trivial results of linear algebra:
\begin{equation}
\gamma_5 \big(\dslash + m + \frac{\mu_I}{2}  \gamma_0\big) \gamma_5 =
\big(-\dslash + m - \frac{\mu_I}{2} \,
 \gamma_0\big)
 =  (\dslash + m - \frac{\mu_I}{2}\,  \gamma_0)^{\dagger}\; .
\label{gammaprop}
\end{equation}
The final equality is based on  the fact that $\dslash$ is
anti-Hermitian (in Euclidean space) but the unit operator and
$\gamma_0$ are each Hermitian. The cyclic property of the
determinant means that one can express the second functional
determinant in Eq.~(\ref{ZI}) as $\, {\rm det}(\dslash + m +
\frac{\mu_I}{2}\, \gamma_0) = \, {\rm det}\left(\gamma_5 (\dslash
+ m + \frac{\mu_I}{2}\, \gamma_0)
 \gamma_5 \right )$. This fact along with
Eq.~(\ref{gammaprop}) gives
\begin{equation}
{\rm det}\big(\dslash + m + \frac{\mu_I}{2}\,  \gamma_0\big) =\left[ \,
{\rm det}\big(\dslash + m - \frac{\mu_I}{2}\, \gamma_0\big) \right
]^*\label{CC} \, .
\end{equation}
Using Eq. (7) along with Eq.~(\ref{ZI}) allows one to deduce that
\begin{equation}
Z_I(T,\mu_I) = \int  d [A] \left | \, {\rm det} \big( \dslash + m -
\frac{\mu_I}{2}  \, \gamma_0 \,\big ) \, \right |^2 e^{-S_{YM}} \; \;
. \label{ZI2}
\end{equation}
Note that this  expression is of the same form as the right-hand
side of Eq.~(\ref{partitionineq}). Thus we see that inequality
(\ref{partitionineq}) together with Eq.~(\ref{ZI2}) yields a
useful inequality,
\begin{equation}
 Z_I\Big(T,\,\frac{2 \mu_B}{N_c} \Big) \ge Z_B(T,\mu_B) \; .
 \end{equation}
This inequality for the partition functions along with the
standard relationship of the free energy to the partition function
implies that
\begin{equation}
{ G}_B(T,\mu_B) \ge { G}_I\Big(T,\, \frac{2 \mu_B}{ N_c}\Big) \; .
\label{ineq1}
\end{equation}
Inequality (\ref{ineq1}) is the principal result of this section.

As discussed above, although the results in this paper are derived
for two flavor QCD, they can be generalized trivially. It should
be immediately clear that the argument goes through without change
for QCD with two degenerate light flavors and  any number of
additional heavy flavors.  The only modification is that the
chemical potential term must be understood as being the chemical
potential  associated with  the up and down quarks and not the
full baryon chemical potential. The reason it goes through is
straightforward.  The various functional integrals in this more
general case include functional determinants for the heavy
flavors.  However, since, as noted above, the chemical potential
only affects the light flavors, these additional functional
determinants are real and non-negative.  Because the inequalities
depend only on the fact that various terms in the functional
integrals are real and positive, the presence of these extra
functional determinants do not alter the preceding inequalities.
It is  worth noting that this more general case is significant:
in nature QCD has two light quarks which are nearly degenerate
and have additional heavy flavors.

There is another scenario in which the inequalities hold.  Suppose
one considers the general case and looks at the full baryon
chemical potential ({\it i.e.,} a chemical potential coupled to
all flavors of quarks) . Suppose further that one is working in a
regime in which the $\overline{s} \gamma_0 s = \overline{c}
\gamma_0 c =\overline{b} \gamma_0 b = \overline{t} \gamma_0 t
=0$.  In such a regime, the total baryon  number in fact comes
from up and down quarks so the previous derivation holds. It
should be noted that such a regime actually occurs.  In
particular it happens at zero temperature if one works below the
critical chemical potential for strangeness condensation.

It is useful to consider how inequality (\ref{ineq1}) may prove
useful. Recall that standard Monte Carlo methods fail for QCD at
finite baryon chemical potential and low temperatures. Moreover,
it is generally believed that weak coupling techniques valid at
very high densities which lead to nonperturbative phenomena in a
manner very similar to conventional BCS theory\cite{Son,rho} are
thought to work only at extraordinarily high densities.  Thus for
the foreseeable future all studies of phenomenological
significance for relatively cold dense matter will of necessity
by based on simplified models \cite{model,rho} rather than  QCD.
There is nothing wrong with using simplified models; virtually all
of the theory of traditional nuclear physics has been made from
the perspective of simplified models and not QCD.  On the other
hand, models need to be constrained in order to be useful.
Empirical data is one way to constrain model building.  To the
extent possible, though, one ought to constrain models directly
from QCD. Inequality (\ref{ineq1}) may prove very useful for this
purpose. While one may have to model the left-hand side of the
inequality, the right-hand side is amenable to lattice QCD
simulations.  The reason for this is precisely the same reason
the inequality was derivable in the first place: namely, the
integrand of the functional integral for $Z_I$  is manifestly
real and non-negative \cite{AlfKapWil99}.  Indeed preliminary
lattice studies have been done for this quantity both for
quenched QCD \cite{Kog1} and for full QCD \cite{Kog2}. These
studies have been done on rather small lattices and it is not
clear just how reliable they are. However, lattice calculations
will undoubtedly improve with time and eventually may provide
important constraints on model building through inequality
(\ref{ineq1}).

\section{A Bound on the Nucleon Mass\label{sec:ineq2}}

The topic of this section is off of the main line discussed in
this article.  However, it is worth pursuing since a significant
result for the nucleon mass emerges naturally as a corollary to
inequality (\ref{ineq1}).  As will be seen, the bound is not very
stringent, but it is still of interest because it is a direct
result of QCD. Moreover, it provides a solution to a very old
problem. Nussinov originally derived a bound in the context of
QCD-inspired models: the nucleon mass must be greater than or
equal to $\frac{3 m_\pi}{2}$\cite{Nus83}. Weingarten, in his
seminal paper introducing QCD sum rules, attempted to place a
bound on the nucleon as being larger than some multiple of the
pion mass\cite{Wei83}. This attempt, unlike that of Nussinov,
directly used QCD. However, the attempt failed---the method was
only valid for QCD in a world of six or more degenerate light
flavors. But this certainly does not correspond to the real world.
Weingarten also suggested an alternative approach which did not
require six degenerate flavors. However, this method depended on
plausible but unproved assumptions about the behavior of the
quark propagator in background gauge potentials.  Nussinov and
Sathiapalan \cite{NusSat85}  were able to derive a QCD-based bound
that $M_N
> \frac{N_c \, m_\pi}{2} $.  Their derivation did not rely on {\it ad
hoc} assumptions about the quark propagator, and it holds for two
degenerate flavors. However, the derivation only works in the
large $N_c$ limit of QCD. Therefore, prior to ref.~\cite{TDC1}
there were no known rigorous bounds on the nucleon mass from QCD.

One particular remarkable fact about the bound on the nucleon mass
obtained in  ref.~\cite{TDC1} is that it is based on thermodynamic
arguments.  The usual way masses are bounded in QCD inequalities
is via the study of Euclidean space correlation functions.

The derivation begins with the observation that  inequality
(\ref{ineq1}) is valid at any temperature and thus applies at
$T=0$.  The zero temperature system has remarkably simple
thermodynamic properties: the system is in a single quantum state
(that is to say, thermal fluctuations are completely absent).  The
quantum state is simply the one that minimizes the $G=H-\mu N$,
where $H$ is the Hamiltonian, $G$ is the appropriate free energy
and $\mu$ is the appropriate chemical potential (either isospin or
baryon); $N=V_3 \rho$ is the associated particle number. The
chemical potential serves to select the quantum state by altering
the free energies of  the various quantum states according to the
particle number.  Assuming the spectrum has a gap, an increase in
the chemical potential from zero (at zero temperature)  will do
no nothing until it reaches a critical value where the free
energy of a quantum state other than the vacuum drops below the
vacuum state.  Below this critical value the density must be zero
at zero temperature.  Note that it is precisely due to the
existence of such a critical chemical potential that the Silver
Blaze problem arises.

The critical chemical potentials can be defined by the following
relations:
\begin{eqnarray}
{ G}_B(T=0,\mu_B) &  = &  0\;
\;{\rm for} \; \;|\mu_B| < \mu_B^c \; ,\nonumber \\[1mm]
{ G}_B(T=0,\mu_B) &  < & 0\;
 \; {\rm for} \; \; |\mu_B| > \mu_B^c \; , \nonumber \\[1mm]
{ G}_I(T=0,\mu_I) &  = & 0\;
\; {\rm for} \; \; |\mu_I| < \mu_I^c \; ,\nonumber \\[1mm]
{ G}_I(T=0,\mu_I) &  < & 0 \;
 \; {\rm for} \; \; |\mu_I| > \mu_I^c \;.
\label{crit}
\end{eqnarray}
Inequality (\ref{ineq1}) along with the relations defining the
critical chemical potential (\ref{crit}) imply that
\begin{equation}
\mu_B^c \ge \frac{N_c \, \mu_I^c}{2} \label{ineq2} \;.
\end{equation}

Now inequality (\ref{ineq2}) is specified in terms of a critical
chemical potential and we want a relation on the nucleon mass.
How can we relate the two?  The answer is straightforward:
$\mu_B^c$ is bounded from above by the nucleon mass.  This can be
seen rather trivially from a variational argument.  Focus on a
quantum state that we know is an eigenstate of the Hamiltonian: a
single nucleon at rest. The free energy of this state is known.
The energy is $M_N$ while the baryon number is unity, thus the
free energy is $G_B = M_N - \mu_B$.  It is obvious that the free
energy of this state is less than zero when $\mu_B \ge M_N$. This
means that there exists a state of lower free energy than the
vacuum whenever $\mu_B \ge M_N$.  Of course, it is logically
possible that there are states of lower free energy than the
vacuum for some value $\mu_B$ less than the nucleon mass implying
a critical chemical potential of less than $M_N$. Indeed, that is
what happens in nature. Extrapolations of the masses and
densities of finite nuclei (while removing Coulomb effects) to an
infinite system \cite{nucmat} lead to  the conclusion that in the
absence of Coulomb effects, bound infinite nuclear matter forms.
Since it is bound the energy per particle is less than that of
isolated nucleons.  The transition to infinite nuclear matter is
first order at zero temperature; just below $\mu_B^c$ the system
has zero energy and zero density, while just above the system has
nonzero energy and nonzero density. Thus $\mu_B^c = M_N - B$
where $B$ is the binding energy per nucleon of infinite nuclear
matter. For the present purpose the essential observation is that
regardless of whether $\mu_B^c$ is equal to or less than $M_N$,
it cannot be greater:
\begin{equation}
\mu_B^c \le M_N \label{critbineq}\;.
\end{equation}
Inequalities (\ref{ineq2}) and (\ref{critbineq}) together yield a
bound on the nucleon mass,
\begin{equation}
M_N \ge \frac{N_c \,  \mu_I^c}{2} \; . \label{ineq3}
\end{equation}
Inequality (\ref{ineq3}) is a principal result of this section.
We have succeeded in bounding the nucleon mass from below by
another physical observable.

As written, inequality (\ref{ineq3}) is of limited use.  We have
no direct way  to measure or compute $\mu_I^c$ without further
assumptions, although  we do have strong theoretical grounds for
believing that $\mu_I^c= m_\pi$. We can, however, turn the
inequality around to make a rigorous statement which can be
checked.  Recall that $\mu_I^c$ is the energy per unit isospin of
the state in QCD with the lowest energy per unit isospin.  We can
name this state $X$ and denote its mass $m_X$ and  isospin $I_X$
so that $\mu_I^c= \frac{ m_X}{ I_X }$. Inequality (\ref{ineq3})
can then be written as
\begin{equation}
M_N \ge \frac{N_c \, m_X}{ 2 \, I_X }  \; . \label{ineq4}
\end{equation}
where $X$ is some state which exists in QCD.  This is a sharp
prediction of QCD that can be checked. Taking $X$ to be the pion
we see that the inequality is  satisfied by more than a factor of
4.

\section{The Silver Blaze Problem}
\subsection{The Isospin Silver Blaze Problem}
Let us now turn to the Silver Blaze problem.  For simplicity of
presentation we consider QCD with two degenerate light flavors.
In this section we will consider  the theory at zero temperature
and a nonzero but small chemical potential (either for baryon
number or isospin or a combination thereof).  For concreteness let
us start the discussion for the case of an isospin chemical
potential (at zero baryon chemical potential). While this problem
is not particularly interesting phenomenologically, it raises
many general questions which have analogs in the more general
case and has the virtue of being solved \cite{TDC2}.

Phenomenologically this system is well understood  at low chemical
potential \cite{ss}.  The system remains in the vacuum state with
zero energy density and isospin density for all $|\mu_I|$ less
than  $m_\pi$, which serves as the critical point.  At the
critical point there is a second-order phase transition.  The
state above the transition is a pion condensate.   It is very easy
to explain this behavior in terms terms of eigenstates of QCD.
The $\mu_I^c$---the critical value of $\mu_I$---is simply the
energy per unit isospin for the state of the system with the
smallest energy per unit isospin. For this system it is a pion at
rest.  However, while this interpretation is trivial the
connection to the QCD lagrangian remains quite obscure; we have no
 simple way to obtain the eigenstates starting directly from QCD.

The point of the present study is to try to understand what is
going on in terms  of a Euclidean space functional integral
formulation of the theory.  The reason for doing this is
twofold.  In the first place, Euclidean space functional
integrals are a general, powerful, theoretical tool.  Secondly,
lattice QCD is formulated in terms of them.

The zero temperature limit introduces subtleties.  Thus it is
simpler to work at finite (but small) temperature  at the outset
and then consider the limit as $T \rightarrow \infty$ at an
appropriate later stage. The key quantity of interest is the free
energy.  It is given by $G_I(T, \mu_I)= E - T S - \mu_I I_3$ (where $E$,
$T$, $S$ and $\mu_I$ are the energy, temperature, entropy
density, isospin and chemical potential, respectively). To help
keep things well defined, we work in a finite (but large) box
with a volume denoted by  $V$. The infinite volume
(thermodynamic) limit V can be taken at the end of the day. In
this limit it is natural to express results in terms of intensive
quantities such  as the energy density, the free energy  or
isospin density.  The free energy is related to the grand
partition function in the standard way: $Z_I(T \mu_I)= e^{-\beta
G(T, \mu_I)}$ (where $\beta =1/T$).  As seen in Subsect.\,\ref{ssq} the grand partition function can be represented as the
following functional integral,
\begin{equation}
Z_I(T,\mu_I) \, = \int  d [A] \,  \, \left | {\rm det} \left (
\dslash + m - \frac{\mu_I \gamma_0}{2} \, \right ) \right |^2
e^{-S_{YM}} \;. \label{funcint}
\end{equation}

The essential issue is how the chemical potential influences the
free energy.  From the functional integral expression it is clear
that the chemical potential influences the free energy through the
functional determinant of the Dirac operator and only through the
functional determinant. We do not know too much about this
determinant. But one thing we do know is that the determinant is
simply the product of the eigenvalues of the Dirac operator $ {\rm
det}\left( \dslash + m - \frac{\mu_I \gamma_0}{2} \,
 \,  \right )   =   \prod_j \lambda_j \; \; \;
 {\rm where} \; \; \; \left( \dslash + m - \frac{\mu_I \gamma_0}{2} \,
 \,  \right ) \psi_j  =  \lambda_j \psi_j$ .

This is at the crux of the Silver Blaze problem.  If we knew
nothing else, it would be natural to assume that for any given
gauge field configuration, the eigenspectrum of the Dirac operator
with $\mu_I =0$  differs from the eigenspectrum with any nonzero
$\mu_I$ .  Indeed, it is  naturally to expect that {\it every
eigenvalue} is would be different.  The reason for such an
expectation is simply the lack of any known  reason why the
eigenvalues should not depend on $\mu$.  Again, in the absence of
any other knowledge, one would also naturally assume then that
{\it for every gauge} configuration, the functional determinant
with nonzero $\mu_I$ would differ from  the functional determinant
$\mu_I=0$. Since all functional determinants appear to depend on
$\mu_I$ it is also natural to conclude that $Z_I(T,\mu_I)$ must
depend on $\mu_I$ for any nonzero $\mu_I$.  Nothing about this
expectation seems to depend in a critical way on the temperature;
it would seem to hold at $T=0$. Thus one has a natural
expectation that at $T=0$ any nonzero chemical potential would
alter the free energy. Obviously this expectation is completely
wrong.  At $T=0$ the free energy is exactly equal to its vacuum
value whenever $|\mu_I| < m_\pi$.   The Silver Blaze problem is
about how to understand the ``curious incident'' of nothing
happening to the free energy  for the entire regime $|\mu_I| <
m_\pi$ in the context of a functional integral treatment.

The  insight needed for the solution of the isospin Silver Blaze
problem is that instead of focusing on the eigenspectrum of the
Dirac operator, $\dslash +m $, one should instead focus on the
eigenspectrum of the $\gamma_0$ times the Dirac operator. Why is
this of interest?  To begin with, note that product rule for
determinants implies that
\begin{equation}
{\rm det}\!\left (\! \dslash+ m - \frac{\mu_I  \gamma_0}{2}
 \right )\!  = \frac{ {\rm det}\Big(\!\gamma_0 \left( \dslash+ m - \frac{\mu_I  \gamma_0}{2}
 \right )\! \Big ) } {{\rm det}\left (\gamma_0 \right)}= {\rm det}\Big (\gamma_0 \left( \dslash+ m - \frac{\mu_I  \gamma_0}{2}
 \right )\! \Big )
\end{equation}
where the last equality exploits the fact that ${\rm det}\left
(\gamma_0 \right)=1$.  Actually this is a bit of a swindle since
the matrices are infinite but it indicates why $\gamma_0$ times
the Dirac matrix may be of interest. A more legitimate way to
express things can be found using some simple linear algebra:
\begin{equation}
{\rm det}\!\left ( \dslash+ m - \frac{\mu_I  \gamma_0}{2}
 \right )\!  =
{\rm det}\left ( \dslash + m  \, \right ) \, \exp\left\{- \frac{1}{2}
\int_0^{\mu_I} \!\!\!\! {\rm d} \mu_I'\, {\rm tr}\, \frac{1}{ \gamma_0
(\dslash + m) -  \frac{\mu_I'}{2}} \right\}  \; .
 \label{detform}
\end{equation}
This indicates that a knowledge of the eigenvalues of $\gamma_0$
times the Dirac operator at various values of the chemical
potential will enable one to find the relevant trace and to do the
integral to find the determinant of interest.

The details of how to characterize these eigenstates and compute
the determinant are given in detail in Refs.~\cite{TDC2,Adams}.
Most of these technical details are omitted here but a few of the
salient results will be noted.

The first important result is that the anti-periodic boundary
conditions on the eigenstates  in the trace along with the
hermiticity properties of the various operators imply that the
eigenfunctions of $ \gamma_0 (\dslash + m) -
\frac{\mu_I'}{2}  $ are arranged into families which can be
denoted by two indices; an index $j$ representing an ``intrinsic''
eigenstate, and an index $n$ representing a phase factor
indicating which anti-periodic solution one is studying:
\begin{equation}
 |\psi_{j n+1} \rangle  =  e^{\frac{i 2 \pi  t}{\beta}}|\psi_{j n}  \rangle \,,\qquad
\lambda_{j n}  =  \epsilon_j - \frac{\mu_I'}{2} + i \left(
\frac{\phi_j}{\beta} + \frac{(2 n +1 ) \pi}{\beta} \right ) \; .
\label{eps}
\end{equation}
The eigenvalue of the operator $\lambda_{j n}$ has  a real and an
imaginary part. To uniquely specify this decomposition a
condition on the phase needs to be imposed.  Here we take the
condition that $-\pi \le \phi_j < \pi$.

The form of Eq.~(\ref{eps}) should look familiar. Apart from the
phase factor $\phi_j$ it is of the same form as for  free
noninteracting fermions.  Of course, in the case of a
noninteracting particle, $\epsilon_j$ has a simple interpretation:
it is the energy of a mode.  Thus we will denote $\epsilon_j$
(the real part of eigenvalue) as a quasi-energy. These
quasi-energies are fundamentally different from  energies in some
essential ways.  In the first place they depend on the background
gauge field configuration.  Moreover we have no analytic
expressions for them.

The trace can be done in two parts: summing over the
quasi-energies, and parameterizing the imaginary parts
parameterized by the index $n$.  The sum over $n$ is the analog
of a typical Matsubura sum\cite{Matsubura} and can be done
explicitly. Straightforward algebra then yields
\begin{eqnarray}
&{}&\frac{{\rm det}\left ( \dslash+ m - \frac{\mu_I  \gamma_0}{2}
\, \right ) }{ {\rm det}\left ( \dslash + m  \, \right )} =
\exp \Big( - i \sum_j \phi_j \, \theta (\epsilon_j)
  \theta (|\mu_I| -    2 \epsilon_j) \Big ) \nonumber \\[1mm]
&{}&   \, \times \,
 \exp \Big(\, \frac{\beta }{2} \sum_j  \theta (\epsilon_j)  \,
 \theta (|\mu_I| - 2 \epsilon_j ) \, \left(|\mu_I| -2 \epsilon_j \right )
 + {  O}\big(e^{- \beta \Lambda}\big) \Big )
 \label{finaldetform}
\end{eqnarray}
where the fact that we are ultimately interested in the zero
temperature limit has been used to  obtain $\theta$ functions
from hyperbolic tangents which emerge from the Matsubura sum. At
this stage we will take the zero temperature limit and neglect
the exponentially suppressed terms.  Note, however, that this is
making the assumption that the quantity of interest, the free
energy at nonzero chemical potential, is smooth in the zero
temperature limit.

The theta functions in Eq. (\ref{finaldetform}) gives an obvious
hint as to how the isospin Silver Blaze problem may be solved.
They imply that at zero temperature the functional determinant
for any gauge configuration is precisely equal to its $\mu_I=0$
value unless $\frac{|\mu_I|}{2}$ is greater than the quasi-energy
of the minimum positive quasi-energy mode.  Thus, if there is a
gap in the quasi-energy spectrum for a given field configuration,
then at least for that configuration nothing happens until the
gap is reached.

However, this is not sufficient to resolve the Silver Blaze
problem by itself.  A full resolution requires that a gap in the
spectrum exists  for the configurations that contribute with
nonzero weight to the functional integral at zero temperature.
One needs a formal way to specify this and it is naturally given
in terms of a spectral density: ${\hat\rho}(\epsilon)$,
\begin{equation}
{\hat\rho}(\epsilon) \equiv \sum_j \delta(\epsilon-\epsilon_j)  \;
\label{specden}
\end{equation}
where $\epsilon_j$  is the $j^{\rm th}$ quasi-energy  (for a given
configuration).    Using the Matsubura sum and generic properties
of the propagator, a very useful expression is obtained for the
spectral density \cite{TDC2}
\begin{equation}
{\hat\rho}(\epsilon)  = \frac{1}{2 \beta}\,
\frac{\partial}{\partial \epsilon} \,{\rm tr} \left [   \left(
\gamma_0 (\dslash + m) -  \epsilon \right)^{-1} + \left( (-
\dslash + m)\gamma_0  -  \epsilon \right)^{-1} \right] + {
O}\big(e^{- \beta \Lambda}\big) \; .
 \label{specden2}
\end{equation}
Next let us introduce a notation to indicate averaging over gauge
configurations:
\begin{equation}
\langle \hat{{ O}} \rangle_{T, \mu_I} =  \frac{1}{ Z_I(T,\mu_I)}
\, \int  d [A] \, \hat{{ O}} \, \left | {\rm det} \left ( \dslash
+ m - \frac{\mu_I \gamma_0}{2} \, \right ) \right |^2 e^{-S_{YM}}
\label{av} \; . \end{equation}
 We can define the minimum relevant
 positive quasi-energy, $\epsilon_{\rm min}$: $\langle \hat{\rho}(\epsilon)
\rangle_{0 , 0} = 0 $ if and only if $|\epsilon | < \epsilon_{\rm
min} $.

With these notational preliminaries in place, the isospin Silver
Blaze problem is resolved provided that two conditions are
satisfied:
\begin{eqnarray}
i) &{}& \; \; \left\langle \hat{\rho}(\epsilon) \right\rangle_{0, \mu_I} =
\left\langle \hat{\rho}(\epsilon)\right\rangle_{0, 0} \; \; {\rm for \; all } \; \; \mu_I < 2 \epsilon_{\rm min}\,, \nonumber \\[2mm]
ii) &{}&\epsilon_{\rm min} = \frac{m_\pi}{2} \; .  \nonumber
 \end{eqnarray}
It is easy to see that these two conditions do indeed resolve the
problem. If one uses the relation of the free energy in terms of
$Z_I$ along with Eqs.~(\ref{funcint}) and (\ref{specden}) then
\begin{equation}
\frac{\partial G(0,\mu_I) }{\partial \mu} =  2
\int_0^{\frac{\mu_I}{2}}\! {\rm d} \epsilon  \left\langle \hat{\rho}
(\epsilon)\right\rangle_{ 0,\mu_I} \; .
\end{equation}
This means that at zero temperature  and $|\mu_I| < m_\pi$ and if
these two conditions are true,  the free energy will be
independent of $\mu_I$  and thus equal to  its vacuum value. This
in turn means  the expectation value of the isospin vanishes.

The validity of these two conditions can be established given one
basically innocuous assumption---that there is no first order
phase transition for $T=0$ and $|\mu_I| < m_\pi$.  This
assumption is highly plausible from first principles and  is
known to be true in nature.  Condition i) can then be established
using straightforward methods which are detailed in
ref.~\cite{TDC2}. Condition ii) is a bit more interesting.  The
trick is to study the charged pseudoscalar susceptibility
$\chi_{\rm ps}^+ = \int {\rm d}^4 x \langle J_{-}(x) J_{+}(0)
\rangle$ (with $J_{+}=\overline{d} \gamma_5 u$).  The key point
is that it is expressible as a functional integral;  using
similar techniques to those discussed above, one obtains
\begin{equation}
 \chi_{\rm ps}^+(T, \mu_I)
= \frac{1}{ V}\int {\rm d}\epsilon \, \,  \frac{\left\langle \hat{
\rho}(\epsilon)\right\rangle_{T , \mu_I} \left ( 1+ { O}\big(e^{-\beta \Lambda}\big)
\right)}{
    \left | 2\,\epsilon  - \mu_I  \right| } \;.
    \label{chif}
\end{equation}
Thus $\chi_{\rm ps}^+$ will diverge when $\frac{\mu_I}{2}$
reaches the smallest value of $\epsilon$ for which
$\langle\hat{\rho}(\epsilon)\rangle_{ 0, \mu_I}$ is nonzero.
Condition i) then implies that this occurs at $\epsilon_{\rm
min}$.  Phenomenologically, in the absence of a first order
transition, $\chi_{\rm ps}^+$ diverges in the infrared when the
chemical potential reaches the mass of the lowest excitation with
these quantum numbers, namely, $m_\pi$.  This completes the
demonstration.

\subsection{The Baryon Silver Blaze Problem}

The baryon Silver Blaze problem is far more interesting.  The
problem is how can one use a functional integral formulation to
understand how it happens that for $\mu_B < M_N -B$ (where $B$ is
the binding energy for nucleon of nuclear matter ) the system is
unchanged from its vacuum.  In the first place it is of much
greater phenomenological importance than the isospin case. After
all, it is at the crux of understanding infinite nuclear matter
from QCD. It is also of far more interest theoretically than its
isospin cousin. The isospin Silver Blaze problem is resolved in a
direct way: all of the functional determinants in the
configurations which matter {\it are} unchanged from their vacuum
values.

Can the baryon Silver Blaze problem be resolved in a similar way?
The answer depends on the regime in which one works. First
consider a regime in which $0<\mu_B< 3 m_\pi /2$. In this regime
the derivation given for the isospin Silver Blaze problem applies;
all of the gauge configurations which contribute have a functional
determinant identical to that of the vacuum.   We note that there
is a paradox associated with the baryon Silver Blaze problem which
applies in this regime.   From Sect.\,\ref{sec:ineq1} we see that
${ G}_B\big(T,\mu_B) \ge { G}_I(T, \frac{2 \mu_B}{ N_c}\big)$ and that
this holds at any temperature including $T=0$.  The origin of
this inequality was simply the phase of the functional
determinant.  Naively one would expect this phase factor to
differ from unity for all gauge configurations since all
eigenvalues of the Dirac operator are complex.  This in turn
leads  to an expectation that the inequality should not be
saturated, and that ${ G}_B(T=0,\mu_B) \ge { G}_I\big(T=0, \frac{2
\mu_B}{3}\big)$ for all $\mu_B$.  But in the present regime this does
not happen.  Both $G_B$ and $G_I$ are the vacuum value and, hence,
they are equal.  Why was the expectation wrong? The derivation in
the isospin Silver Blaze problem neatly explains this.

Consider Eq.~(\ref{finaldetform}). Note that there are theta
functions for the quasi-energies contributing to the phase factor.
These are precisely the same theta functions as those for
contributions to the magnitude of the functional determinant.
Thus, the fact that the configurations which matter to the
functional integral have their magnitudes unchanged from their
vacuum value in this regime (the resolution of the isospin Silver
Blaze problem) also implies that the phases of the functional
determinant are unchanged for the relevant configurations.  This
explains why in this regime $G_B = G_I$ rather than being less.

Next let us  turn to the regime $3 m_\pi/2 < \mu_B  <M_N-B$.  In
this case it is clear that the nature of the solution of the
Silver Blaze problem is qualitatively different from the isospin
case. To see this let us again return to Sect.\,\ref{sec:ineq1}.
Recall that the fundamental reason why an inequality was derived
in that case was because the integrand for the free energy with a
baryon chemical potential differs from the integrand for the free
energy with the appropriate isospin chemical potential only due
to a phase factor.  Now in this kinematic regime we know
phenomenologically that ${ G}_B(T=0,\mu_B)$ is at its vacuum value
while ${ G}_I\big(T=0, \frac{2 \mu_B}{3}\big)$ is below the vacuum value
due to pion condensation.  Pion condensation  implies that the
functional determinants  of gauge configurations that contribute
{\it are} altered from their vacuum value.  The only way the
baryon chemical potential can leave the free energy unaltered is
because of the  phase factors.  However, this implies a very
large conspiracy---the entire effect of the magnitude of the
functional determinants increasing must be {\it exactly}
compensated by averaging over the phases.  This phenomena is
clearly qualitatively quite distinct from the behavior
responsible for the isospin Silver Blaze problem.

What is the origin of this conspiracy?  At the present time this
is unknown.  A pessimistic view is that answering this question is
tantamount to solving QCD analytically and, hence, is intractable.
An optimistic view is that the issue may become clear if one can
find a suitable  reorganization of the problem.  After all, the
isospin Silver Blaze problem also looked intractable until it was
realized that the key was to express things in term of the
eigenvalue of $\gamma_0$ times the Dirac operator rather than the
Dirac operator itself.  Where should we look for hints about how
such a reorganization might be accomplished?  Although we do not
really know, there are some obvious sources for inspiration. One
is the region just above $3 m_\pi/2 = \mu_B$.  In this region the
relevant configurations have functional determinants slightly
larger than at $\mu_B=0$ which must be canceled by the phase
effects during averaging. Since the functional determinant is
small one may be able to derive analytic expressions on the
necessary conditions for cancellation which in turn may give a
clue as to how things should be organized.  The second place to
look is in the large $N_c$ limit of QCD.  A diagrammatic analysis
in the large $N_c$ limit suggests that effects linking the
functional determinant for the up quarks with the functional
determinant for the down quarks is suppressed by $1/N_c$.  Thus
one expects that in the large $N_c$ limit the critical chemical
potential for the up quark charge and the down quark charge are
identical.  Clearly this does not happen as it implies the
critical baryon chemical potential is just $N_c/2$ times the
isospin chemical potential which clearly fails in the chiral
limit. Therefore, understanding the breakdown of the large $N_c$
approximation for these quantities may well provide a clue.

\end{document}